\documentclass[twocolumns]{aa}
\usepackage{graphicx}
\usepackage{txfonts}
\begin{document}

   \title{The VLT-VIMOS Mask Preparation Software}

   \author{D. Bottini\inst{1}, B. Garilli\inst{1},
   D. Maccagni\inst{1}, L. Tresse\inst{2},
\and V. Le Brun \inst{2}
\and O. Le F\`evre \inst{2}
\and J.P. Picat \inst{3}
\and R. Scaramella \inst{4}
\and M. Scodeggio \inst{1}
\and G. Vettolani \inst{4}
\and A. Zanichelli \inst{4}
\and C. Adami \inst{2}
\and M. Arnaboldi \inst{5}
\and S. Arnouts \inst{2}
\and S. Bardelli  \inst{6}
\and M. Bolzonella  \inst{7}
\and A. Cappi    \inst{6}
\and S. Charlot \inst{8,10}
\and T. Contini \inst{3}
\and S. Foucaud \inst{1}
\and P. Franzetti \inst{1}
\and L. Guzzo \inst{9}
\and O. Ilbert \inst{2}
\and A. Iovino \inst{9}
\and H.J. McCracken \inst{10,11}
\and B. Marano     \inst{7}
\and C. Marinoni \inst{2}
\and G. Mathez \inst{3}
\and A. Mazure \inst{2}
\and B. Meneux \inst{2}
\and R. Merighi   \inst{6}
\and S. Paltani \inst{2}
\and A. Pollo \inst{9}
\and L. Pozzetti    \inst{6}
\and M. Radovich \inst{5}
\and G. Zamorani \inst{6}
\and E. Zucca    \inst{6}
}

   \offprints{D. Bottini}

   \institute{
IASF-INAF - via Bassini 15, I-20133, Milano, Italy
\and
Laboratoire d'Astropysique de Marseille, UMR 6110 CNRS-Universit\'e de
Provence,  BP8, 13376 Marseille Cedex 12, France
\and
Laboratoire d'Astrophysique de l'Observatoire Midi-Pyr\'en\'ees (UMR
5572) -
14, avenue E. Belin, F31400 Toulouse, France
\and
IRA-INAF - Via Gobetti,101, I-40129, Bologna, Italy
\and
INAF-Osservatorio Astronomico di Capodimonte - Via Moiariello 16, I-80131, Napoli,
Italy
\and
INAF-Osservatorio Astronomico di Bologna - Via Ranzani,1, I-40127, Bologna, Italy
\and
Universit\`a di Bologna, Dipartimento di Astronomia - Via Ranzani,1,
I-40127, Bologna, Italy
\and
Max Planck Institut fur Astrophysik, 85741, Garching, Germany
\and
INAF-Osservatorio Astronomico di Brera - Via Brera 28, Milan,
Italy
\and
Institut d'Astrophysique de Paris, UMR 7095, 98 bis Bvd Arago, 75014
Paris, France
\and
Observatoire de Paris, LERMA, 61 Avenue de l'Observatoire, 75014 Paris,
France
}

   \date{}

   \abstract{
VIMOS (VIsible Multi-Object Spectrograph) is a multi-object imaging spectrograph
installed at the VLT (Very large Telescope) at the ESO (European Southern Observatory)
Paranal Observatory, especially suited for survey work. VIMOS is characterized by its very high
multiplexing factor: it is possible to take up
to 800 spectra with 10 arcsec long slits in a single exposure. To fully exploit its multiplexing potential,
we designed and implemented a dedicated software tool: the VIMOS Mask Preparation Software (VMMPS),
which allows the astronomer to select the objects to be spectroscopically observed,
and provides for automatic slit positioning and slit number maximization within the
instrumental constraints. The output of VMMPS is used to manufacture the slit masks to be mounted in the
instrument for spectroscopic observations.

   }
   \authorrunning{Bottini at al.}
   \maketitle

\section{Introduction}

VIMOS (VIsible Multi-Object Spectrograph) is a survey multi-object spectrograph operational at one
of the Nasmyth foci of the Melipal telescope of
the ESO (European Southern Observatory) VLT (Very large Telescope)
since April 2003. VIMOS was designed and built by the French-Italian VIRMOS Consortium (\cite{Lefevre00}). 
VIMOS is based on a classical focal reducer design replicated
in 4 separate channels operating simultaneously (see Fig. 1). It has 3 operational modes: imaging, with broad
band UBVRIz filters; MOS (Multi-Object Spectroscopy), with a choice of 5 grisms giving
spectral resolutions from 200 to 2500 in the wavelength range 0.37-1 $\mu m$;
and IFS (Integral Field Spectroscopy), whereby 6400 fibers coupled with micro-lenses produce
spectra of a contiguous area of $54\times 54$ square arcsec.

In MOS mode, observers design four slit masks (one for each quadrant) from images taken prior to the
spectroscopy observing run, heheafter called "pre-images", and a user-supplied source catalogue.
Then the masks are manufactured by the MMU (Mask Manufacturing Unit) which is a dedicated laser
cutting machine (\cite{Conti01}).
Once inserted in the focal plane, masks are aligned on the sky before the exposure.

One of the main drivers in the design of VIMOS was the multiplexing factor, i.e. the possibility
of packing as many spectra as possible on the detectors.
For example, at the limiting magnitude $I<24$, with a low-resolution grism, we have been able
to obtain up to 1000 spectra.

The number of spectra that can be packed on a detector depends on the detector size, the length
of the spectra, the length of the slits, the possibility of placing slits over the whole field of view
without losing wavelength coverage, and, last but not least, the spatial distribution of the targets.
To maximize the number of slits,
we have adopted a similar approach to that successfully used by the
Canada-France Redshift Survey (CFRS) on the
MOS-SIS spectrograph at CFHT (\cite{Lefevre94}), allowing
zero order and second order spectra to overlap with
the first order spectra from other slits. This makes for an
efficient packing of spectra on the detectors, with several super-imposed
banks of spectra, and with
strict constraints on slit mask design and data processing
(\cite{Lefevre95}).

To improve on this concept, the VIMOS optical design is
such that slits can be placed anywhere in the
field of view, and the optical cameras image the
spectra on $2048\times 4096$ pixels$^2$, the $4096$ pixels being placed
along the dispersion direction, while the field of view on the
sky is imaged only on $2048\times 2340$ pixels$^2$.
This means that for any slit placed in the field of view,
the full wavelength range of a low resolution spectrum is
recorded on the detector. This property also applies to
higher resolution $R\sim2500$ spectra, which are
fully recorded on the detector for most slit positions in the field of view.

The large number of slits (up to $\sim$200 per quadrant) involved
in the VIMOS spectroscopic observations together with slit positioning constraints
makes it practically impossible for the astronomer to manually choose and place
all the slits, in particular if one wants to maximize the number of slits.
The software tool we are describing in this paper, called VMMPS, allows to design the slit masks
taking into account all the instrument constraints and solves the problem of maximizing the multiplexing factor
given a certain target field (\cite{Garilli99}). This tool has been delivered by our Consortium to ESO, and 
is now distributed by ESO to all successful proposers.

VMMPS provides the astronomer with tools for the selection of the
objects to be spectroscopically observed. It includes interactive object
selection, it handles curved slits, and an algorithm for automatic
slit positioning which derives the most effective solution in terms of
number of objects selected per field. The slit positioning algorithm
takes into account both the initial list of user's targets, and the constraints
imposed either by the instrument
characteristics and/or by the requirement of easier reduceable
data. The number of possible slit combinations in a field is in any
case very high, and the task of slit maximization cannot
be solved through a purely combinational approach. We have introduced
an innovative approach, based on the analysis of the function $N_{slit}/W_{col}$, where
$N_{slit}$ is the number of slits within a column and $W_{col}$ the width of the column.
The algorithm has been fully tested and validated.\\

In this paper we describe the MOS observation preparation flow (Section 2),
the imaging and catalogue handling (Section 3),
the instrumental and data reduction requirements (Section 4) and the slit
positioning optimization code (Section 5). We conclude in Section 6.

\section{The MOS observations preparation flow}

Pre-imaging with VIMOS is required for subsequent spectroscopic follow-up,
even when targets come from a pre-defined catalog.
Pre-imaging is necessary to set the proper coordinate transformation between
an existing catalog with reference astrometry, and the VIMOS coordinate system.
Since mask preparation can be done using pre-established astrometric catalogs,
pre-imaging is not required to be as deep as to detect all the spectroscopic
targets. It should however be deep enough to provide enough targets for a
cross-correlation between the catalog extracted from the image itself and the pre-defined
astrometric catalog.
In the case of a pre-established astrometric catalog, the first step to be performed
by VMMPS is the cross-correlation and the subsequent re-coordination to
the coordinate grid defined by
the VIMOS image. If the catalogue to be used has been obtained direcly from a VIMOS image
this first operation is not needed.
The following step is to define the catalogue "special" objects. These are the objects to be
used for mask alignment (reference), the objects which must be observed in any case (compulsory),
and the objects you do not want to observe (forbidden). Eventually "special" slits, tilted or curved,
can be defined interactively.
The last step is to define the set of objects to be observed. This step is
performed by SPOC (Slit Positioning Optimization Code), that is the core of VMMPS.
SPOC places slits maximizing their number in a VIMOS quadrant, taking into account special objects,
special slits and slit positioning constraints.

\section{Imaging and catalogue handling}

We need tools which allow us either
to select objects from an image obtained with VIMOS (or from a catalogue derived
from this image) or from our own catalogue.
In principle, telescope (VLT) and site (Paranal) characteristics make it conceivable
to use slits as narrow as 0.5 arcsec, a possibility which,
in turn, requires an excellent mapping of the projection of the sky
onto the focal plane.
The mapping of the focal plane onto the CCD is part
of the calibration procedure of the instrument, and it
has been tested that the accuracy of this calibration
is better than one pixel (0.2 arcsec) at any point in the field of view.
The mapping of the sky onto the CCD, or the astrometric calibration
of the images, is a more difficult task, where an accuracy of
0.3 arcsec (absolute astrometry) is commonly
considered as an excellent result. Relative astrometry (i.e.
the relative position of objects within the field of view)
is usually much more accurate (0.1 arcsec or less).

To avoid any problem with the astrometry,
it has been decided to place slits only starting from
a ``preparatory image'' acquired with VIMOS. Thus
object coordinates are already in VIMOS pixels and
the computation of slit coordinates requires only
the well known CCD to focal plane calibration.
Users having other information derived
from observations with other instruments must have the possibility to use their catalogues
to select spectroscopic targets. Due to the intrinsic difficulties
in absolute astrometric calibrations, the most reliable
solution is to follow a two steps approach.
First, acquire a preliminary image with a moderately accurate astrometric solution.
From this image, a catalogue of objects with X and Y coordinates in VIMOS pixels
and approximated sky coordinates can be obtained with any detection algorithm.
The approximated coordinates are then matched against the user's own catalogue.
The detection algorithm finds common objects within a given
tolerance between the user's catalogue and the VIMOS pre-image catalogue.

The new coordination is done using RA and DEC from the user's catalogue and
X and Y from the pre-image catalogue for the common objects. The X and Y
VIMOS coordinates for every object of the user's catalogue are also
computed.
The matching and coordination
processes can be iterated until the requested R.M.S is reached.
The algorithm is based on the WCS libraries and the coordination is
done by fitting the parameters of the CO matrix with a polynomial, with the
WCS fitplate function.

In this way, any approximation in the astrometry of the VIMOS image
is washed out by the subsequent transformation, and the errors
are confined within the errors of the user's catalogue internal relative astrometry and
of the final transformation. The latter can be very accurate, provided
that it is computed using a sufficient number of objects. Tests
on real data have shown that 60 to 80 objects in common between the
pre-imaging and the user's catalog give excellent results in terms
of positional accuracy of the solution.
Any possible rigid offset between the solution found and the absolute
pointing will be corrected for at the telescope, with the fine
pointing procedure foreseen (see below):
this can correct for offsets as large as
3 arcseconds.

\section{Instrumental and Data Reduction Requirements}
\subsection{Masks}

   \begin{figure}
   \centering
   \includegraphics[width=9cm]{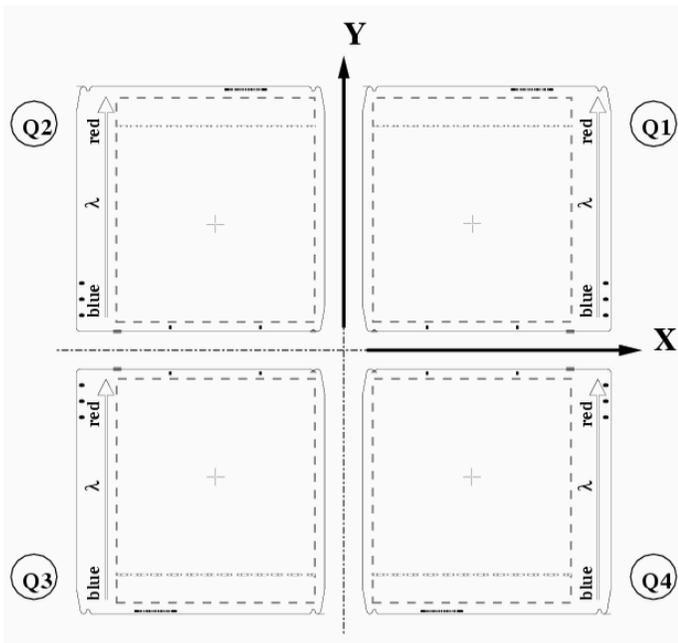}
   \caption{The figure shows (on scale) the location in the 4 VIMOS quadrants
       of the 4 masks with contour and orientation.
       The dispersion direction of the spectra is along the Y axis. The four crosses mark the position of the
       optical axis of each quadrant, while the dashed line indicates the $244\times 279$ mm "useful area" of
       the detector field of view.
   \label{fig1}}
   \end{figure}

The VIMOS focal plane is divided into 4 quadrants, thus 4
masks (1 mask set) are needed for
each MOS observation. The scale at the focal plane where masks must be
inserted is 0.578 mm arcsec$^{-1}$ and
each quadrant has a field of view of about $7\,\arcmin \times 8\,\arcmin$. This
defines the gross dimensions of the masks and, together with the expected
best seeing at the VLT ($\sim\, 0.3\,
\arcsec$), the minimum slit width.

Fig. 1 shows an outline of the VIMOS focal plane with its reference system.
The size of the masks is
$305\times 305$ mm, and the contour geometry is determined by the mechanical
interface with the focal
plane and with the automatic mask insertion mechanism (\cite{Conti01}).

It is important to note that the positioning of masks one
relative to each other cannot be changed, therefore when preparing
masks relative positions of objects must be consistent
over the whole field of view, and not only over a single quadrant.

\subsection{Slit Positioning Constraints}

Within a limiting magnitude of $I=24$, more than 1000 objects are
visible in a VIMOS quadrant.
Not all these objects can be spectroscopically
observed in a single exposure because the overlap of spectra, both in
the spatial and in the dispersion direction, must be avoided.
When low dispersion grisms are used, there can be several spectra along the
dispersion for the same position on the CCD X coordinate (multiplexing).
In this case objects must have a separation greater than their first order
spectrum length (see Fig. 2 case a), and the spectra
resulting from slits placed on objects at different spatial coordinates must not
overlap (see Fig. 2 case b).
Furthermore, low resolution grisms produce second order spectra
which, although of very low intensity ($\sim 3\%$ of the intensity of the first order), can be extremely
annoying for the sky subtraction of the weakest objects. For a better
estimate of the spectral background, it is preferable to have second order
spectra overlapping any eventual first order spectrum in an exact way,
i.e. the two slits placed ``one above the other'' should be of the same
length and aligned (see Fig. case 2 c).

This requirement ensures that the second order sky contamination
of the slit below and the zero order of the slit above (in the
dispersion direction) can be removed together with the first order
sky when the spectral background of a slit is processed.
This process does not
remove the second and zero orders of the objects in the slits
below/above, but it completely eliminates the inconvenience
of having a step function contamination to remove along
the spatial direction of slits when correcting for the sky
background.

On the contrary, VIMOS high dispersion grisms produce first order
spectra approximately as long as
the dispersion direction of a VIMOS quadrant, i.e. only a single column of slits
can be placed along the dispersion direction.

The slit length is set by the object size plus a minimum area of sky
on both side of the object, required for a stable fitting of the sky signal
to be removed during data processing. The size of
this area is suggested to be of the order of 2 arcsec per side (i.e. about 10 pixels)
to ensure a good sky level fitting.

Slits as narrow as 0.5 arcsec are a real
possibility for VIMOS, and it implies an extremely precise pointing and
centering of objects in slits.
The best and safest way to obtain such accuracy is to have
reference objects in the field, to be used to refine the pointing on the sky. Once
the telescope is on the field, an exposure with the mask
inserted is taken, and pointing is refined so that
reference objects fall exactly at the center of
the reference ``holes''. Reference objects must be bright and
point-like, and they must be part of the same user input catalogue.
Reference objects have to be manually chosen. The
slit positioning algorithm has to handle them in a special way, as
they have square holes with a fixed size instead of rectangular slits.
Although in principle only one reference object per quadrant would be required for
pointing refining, a minimum of 2 reference objects per quadrant is recommended.

As reference objects are bright, their second order spectrum contamination is
strong. For this reason, no slit is allowed in the same spatial coordinate range of
reference object holes. The same requirement is applied to curved or tilted slits,
for particularly extended and interesting objects, and no spectrum
overlapping is thus allowed.

These constraints must be fully taken into account by any slit
positioning code.

   \begin{figure}
   \centering
   \includegraphics[angle=-90,width=9cm]{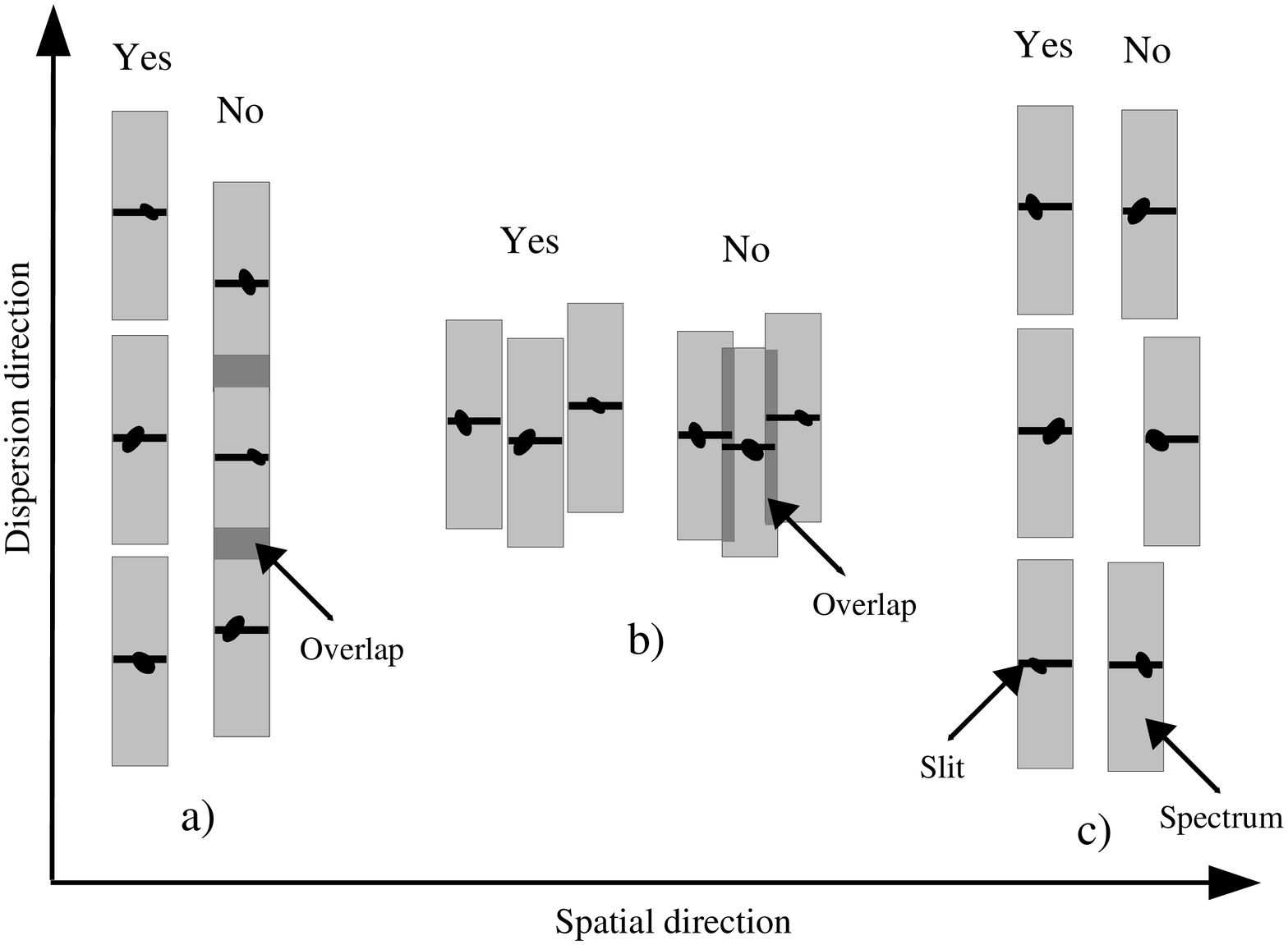}
   \caption{Spatial contraints for MOS slit positioning: "Yes" indicates allowed slit placements,
     "No" indicates forbidden slit placements.Three cases are illustrated (see Section 3.1)
   \label{fig2}}
   \end{figure}

\subsection{Manual selection of objects }

VMMPS is based on the astronomical package SKYCAT distributed by ESO.
SKYCAT has enhanced real time display capabilities and provides basic functions
for catalogue handling. Display and catalogue functions are coupled
by good overlay capabilities and WCS is well supported.
The VMMPS graphical user's interface is a SKYCAT plugin developed in TCL/TK.
Allowed formats are FITS for images and ASCII for catalogues.\\

As slits are positioned with an automated procedure which maximizes the number
of placed slits, there is no way to choose objects ``a priori''.
For all scientific purposes which do not rely on systematic surveys, it
can be a disadvantage. Compulsory objects are the answer to such needs:
the automatic procedure place slits on them (after having
checked that they do not violate the constraints described above)
before anything else, then it tries to place additional slits on objects
randomly chosen from the catalogue.

Forbidden objects are the opposite of compulsory objects: they
are flagged by the user so that no slit is placed on them.

Last, but not least, a number of scientific programs would greatly
benefit from
having curved or tilted slits which better follow the object profile
(gravitational arcs are just the most obvious example). The laser
cutting machine (\cite{Conti01}) is capable of cutting arbitrary
shapes. Therefore, VMMPS allows the user to
interactively design curved or tilted slits.
Curved slits are defined by fitting a Bezier curve to a set of points
chosen interactively by the user.

%______________________________________________________________

\section{Slit Positioning Optimization Code (SPOC)}

The large number of objects and slits involved
in the VIMOS spectroscopic observations together with the positioning
constraints and the particular objects described above,
make it practically impossible for the astronomer to manually choose
and place the slits.
For this reason a tool for the automatic positioning of slits and the maximization
of their number has been implemented: SPOC (Slit Positioning Optimization Code).
SPOC is the core of VMMPS. Given a catalogue of objects,
it maximizes the number of observable objects in a single exposure and
computes the corresponding slit positions.
SPOC places slits on the field of view taking into account
special objects (reference, compulsory, forbidden),
special slits (curved, tilted or user's dimension defined slits)
and slit positioning constraints.
An example of SPOC slit positioning for a VIMOS quadrant is shown in figure 3.

   \begin{figure}
   \centering
   \includegraphics[width=9cm]{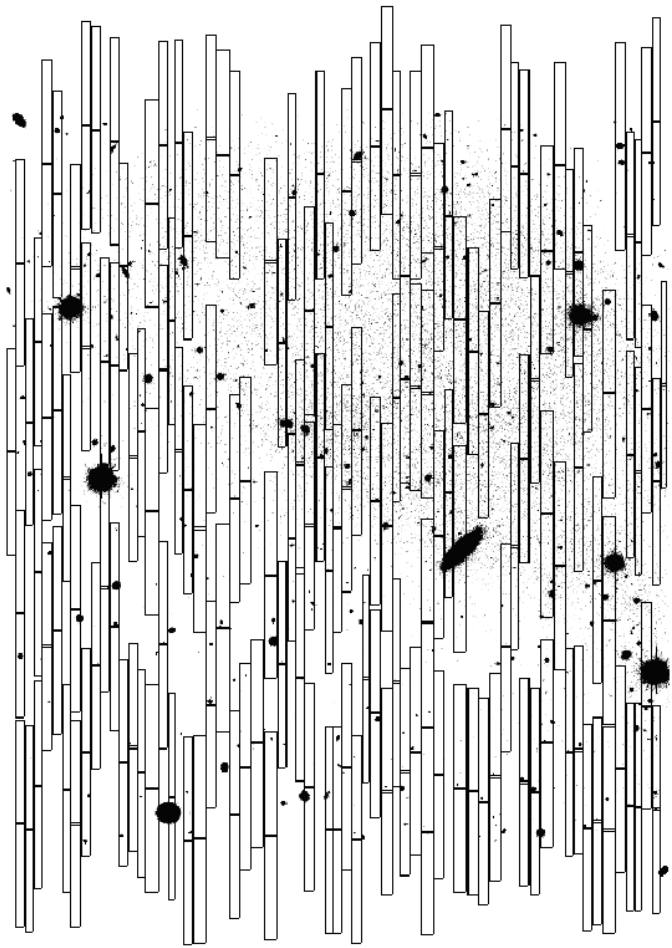}
   \caption{Output of the VMMPS/SPOC algorithm run on a single
  quadrant of a VIMOS image and using an external user catalog:
  the number of slits positioned is maximized by VMMPS using
  all the user and instrument constraints.
   \label{fig3}}
   \end{figure}

The issue to be solved is a combinatory computational problem. Due to the
constraint of slits aligned
along the dispersion direction, the problem can be slightly simplified: the quadrant
area is considered
as a sum of columns which are not necessarily of the same width in the spatial direction.
Slits within the same column have the same length so that the alignment of orders is
fully ensured.
The problem is thus reduced to be mono-dimensional.
It is easy to show that the number of combinations is roughly given
by: $N_c = N_w^{N_{col}}$,where $N_c$ is the number of combinations, $N_w$ the number
of possible column widths, and $N_{col}$ the average number of columns.

The slit length (or column width) can vary from a minimum of {\it 4} arcsec
({\it 20} pixels, i.e. twice the minimum
sky region required for the sky subtraction) to a maximum of {\it 30} arcsec
({\it 150} pixels, limit imposed by the slit
laser cutting machine) in 1 pixel steps. The average number of columns is estimated as the
spatial direction size of the
FOV divided by the most probable slit length, tipically {\it 50}
pixels ({\it 10} arcsec),
i.e. $2048/50 = 41$.
The number of combinations is then: $N_c = (130)^{41}\sim4.7\times10^{86}$,
corresponding to about $10^{60}$ years of CPU work!

The problem is similar to the well-known traveling salesman problem. In the standard
approach, it is solved
by randomly extracting a "reasonable" number of combinations and maximizing over this
subsample.
In our case, due to computational time, the "reasonable" number of combinations cannot
be higher than
$10^8$-$10^9$. So small with respect to the total number of combinations that the result is
not guaranteed to be a good approximation of the real maximum.

Our approach has been to consider only the most probable combinations, i.e. the ones
that have the
highest probability to maximize the solution.
For each spatial coordinate, we vary the column width from the given minimum to the given
maximum, and we count how many objects can be placed in the column. Figure 4 shows
the number of slits in a column divided by the column width ($N_{slit}/W_{col}$) as a
function of the column width.
It is obvious that increasing the column width does not monotonically increase
the number of slits which can be placed in a column. However,
the function has evident maxima, which, of course, depend on the
size of the objects.
For each spatial coordinate, only the column widths corresponding to the five peaks
are worth to be considered,
as they correspond to local maxima of the number of slits per column. The exact
positioning of the peaks
varies for each spatial coordinate, but the shape of the function remains the same.
Using a partition exchange method, the position of the
peaks can be easily found in no more than $6$-$7$ trials.

   \begin{figure}
   \centering
   \includegraphics[angle=-90,width=9cm]{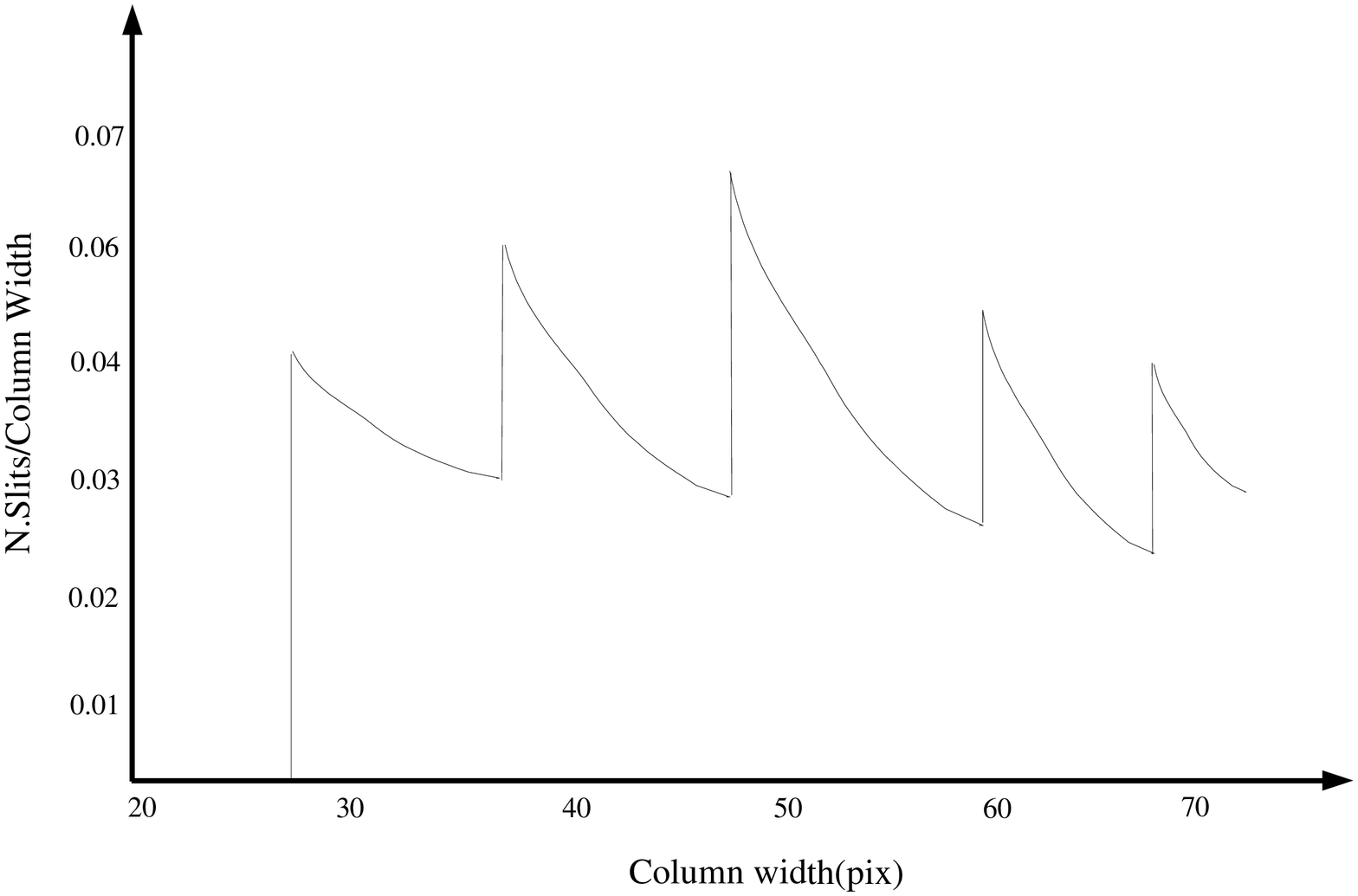}
   \caption[mps_fig4.ps]{Function No.Slits/Column width: increasing the
   column width this function decreases but a maximum is reached every time a new
   slit is placed in the column itself.
   \label{fig4}}
   \end{figure}

For each spatial coordinate we have $K$ (where $K$ is the number of peaks, in our
example $K=4$) possible
columns with their own length and number of slits. Although the number of combinations
to be tested is
decreased (in our example, about $4^{41}=4.8 \times 10^{24}$), it is still too large
in terms of computational time.

A further reduction can be obtained if, instead of simultaneously considering all the columns, we
consider sequentially $M$ subsets of $N$ consecutive columns, which cover the
whole quadrant.
At this point, we should vary $N$ (and consequently $M$) to find the best solution. In practice, when $N$
is higher than $8$-$10$,
nothing changes in terms of number of observable objects. For $N=10$, thus $M=4$
(i.e. $2048/(4\times 50)$), the number
of combinations is reduced to only $4 \times 4^{10}=4 \times 10^6$, that means a few
seconds of CPU work.

As a consequence of the optimization process, small size objects are favoured against
the large size ones. Small objects are statistically fainter than big ones,
so weak objects are slightly favoured against the bright ones and this
can be a drawback for some observational projects.

To overcome this drawback, a less optimized algorithm has also been implemented in SPOC.
This alternative algorithm does not optimize all columns simultaneously, but it builds
the $N_{Slit}/Column_{width}$
function (Fig. 4) column by column without considering object sizes, and it takes only
the maximum.
Then it increases each column width to account for object sizes.

We tested these two different SPOC maximization modalities on a sub-sample of about $90000$ objects
extracted from the VIMOS-VLT Deep Survey database (\cite{Lefevre03}). The magnitude and the radius
distributions of this sub-sample and of the SPOC output (about $7000$ objects) are shown in figure 5.
The excess of small (Fig. 5 case a) and faint (Fig. 5 case c) objects produced by SPOC with
the best optimization modality disappears when the less optimized modality is used (Fig. 5 case b and d) while
the number of placed slits decreases by a few percent.

   \begin{figure}
   \centering
   \includegraphics[width=9cm]{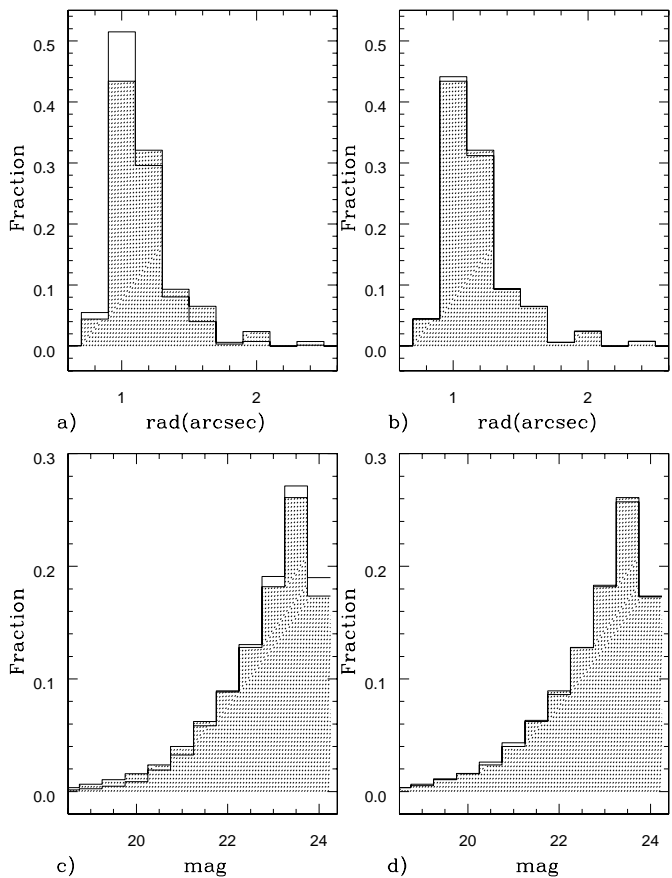}
      \caption{Magnitude and radius distribution: the shaded histograms correspond to the input
       sample, the empty  histograms correspond to the SPOC output.
   \label{fig5}}
   \end{figure}

\section{Conclusions}

\begin{enumerate}
      \item VMMPS provides the VIMOS observer with user-friendly software tools to select the
spectroscopic targets in the intrument field of view, including interactive object
selection, handling of curved slits and maximization of the number of targets once
the input catalogue is defined.
The VMMPS output is a set of files that are passed to the VIRMOS MMU to cut the slits
in the masks and included in the file header of the spectroscopic images for the subsequent data reduction.
      \item VMMPS has been successfully used by the VIMOS team for the preparation of the VIMOS-VLT
Deep Survey (\cite{Lefevre04b}), and by the European astronomical community for the preparation of their specific
observation programs.
      \item VMMPS is a general purpose tool for multi-object spectroscopy, rather than being instrument specific, and
its code has been distributed to GMOS (Gemini) and OSIRIS (Gran Telescopio Canarias).

\end{enumerate}

\begin{acknowledgements}
      The VMMPS has been developed under ESO contract
      50979/INS/97/7569/GWI.
This research has been developed within the framework of the VVDS consortium.
This work has been partially supported by the Italian Ministry (MIUR) grants
COFIN2000 (MM02037133) and COFIN2003 (2003020150).
\end{acknowledgements}

\end{document}